\documentclass[preprint,amsmath,amssymb,aps]{revtex4-1}

\usepackage{graphicx}
\usepackage{dcolumn}
\usepackage{bm}
\usepackage{color, colortbl}

\definecolor{Gray}{gray}{0.9}

\begin{document}

\title{
Principles of efficient chemotactic pursuit}
\author{Claus Metzner}
\email{claus.metzner@gmail.com}
\affiliation{\small 
Department of Physics, Biophysics Group, Friedrich-Alexander University Erlangen, Germany
}
\date{\today}

\begin{abstract}
	
In chemotaxis, cells are modulating their migration patterns in response to concentration gradients of a guiding substance. Immune cells are believed to use such chemotactic sensing for remotely detecting and homing in on pathogens. Considering that an immune cells may encounter a multitude of targets with vastly different migration properties, ranging from immobile to highly mobile, it is not clear which strategies of chemotactic pursuit are simultaneously efficient and versatile. We takle this problem theoretically and define a tunable response function that maps temporal or spatial concentration gradients to migration behavior. The seven free parameters of this response function are optimized numerically with the objective of maximizing search efficiency against a wide spectrum of target cell properties. Finally, we reverse-engineer the best-performing parameter sets to uncover the principles of efficient chemotactic pursuit under different biologically realistic boundary conditions. Remarkably, the numerical optimization rediscovers chemotactic strategies that are well-known in biological systems, such as the gradient-dependent swimming and tumbling modes of E.coli. Some of our results may also be useful for the design of chemotaxis experiments and for the development of algorithms that automatically detect and quantify goal oriented behavior in measured immune cell trajectories.

\begin{description}
\item[Keywords]
Chemotaxis, search strategies, immune cells.
\end{description}
\end{abstract}
\maketitle

\section{Introduction}

Chemotaxis, the ability of cells to detect and follow concentration gradients of specific chemicals, is ubiquitus in biology (For an introduction to the field, see \cite{eisenbach04} and the references therein). It helps sperm cells to find the ovum, directs cell movements during embryogenesis, but also enables organisms to locate food sources and to avoid hostile environments. In particular, chemotaxis plays a vital role in recruiting motile immune cells to sites of infection or to malignant tumors. This recruitment of immune cells is often based on endogenous chemo-attractants, which are released by other host cells that are already at the location where a pathogen has invided the body. However, the fact that individual immune cells are able to find and eliminate tumor cells in a Petri dish, without any assistance, suggests that immune cells can also detect chemical traces emitted by the pathogens themselves. 

In this paper, we are mainly interested in this latter scenario of a single immune cell finding and eliminating several target cells on a two-dimensional plane with periodic boundary conditions. The target cells are modeled as simple agents that move, independently from the immune cell and from each other, with fixed speed and with fixed directional persistence. While migrating, the targets are emitting a chemical substance that acts as a chemo-attractant for the immune cell. This chemo-attractant is assumed to spread quickly within the extracellular medium by linear diffusion. It is also assumed to decay at a constant rate, so that a concentration profile of fixed shape will surround each target cell at any moment.  

The immune cell is modeled as a more complex agent with concentration sensors for the chemo-attractant and with the ability to change its migration behavior accordingly. In the simplest case, the immune cell has only a single chemo-attractant sensor and compares the measured local concentrations between subsequent simulation time steps (temporal sensing). In the more powerful case of spatial sensing, the immune cell uses multiple sensors at different body positions to measure the spatial gradient of the chemo-attractant concentration. 

In order to modulate the migration properties depending on the sensed concentration gradients, the immune cell uses  probabilistic 'stimulus-response functions' with tunable parameters. In the case of temporal sensing, the response function controls the momentary probabilities for being in one of two possible modes of migration, characterized by different speeds and degrees of directional persistence. In the case of spatial sensing,
the response function determines the probability of the immune cell turning clockwise or counter-clockwise. 

The parameters of the response functions are optimized numerically, with the objective to maximize the average number of direct contacts between the immune cell and distinct target cells during a fixed simulation time - a number called the 'search efficiency' $Q$ (Here, we assume that once a direct contact is established, the respective target cell is immediately removed from the system). In order to obtain an immune cell that is not only efficient in finding specific types of targets but also robust against variable target behavior, the simulated immune cell is confronted with a broad spectrum of target cell speeds $v_{tar}$ and directional persistences $\epsilon_{tar}$ during the optimization phase. Once the optimal response parameters are found, we also evaluate the specific performance $Q=Q(v_{tar},\epsilon_{tar})$ of the immune cell as a function of the target cell's migrational properties.

\section{Methods}

\subsection{Cell migration model}
We consider a single immune cell (with index $c=0$) and several target cells (with indices $c=1\ldots N_{tar}$) on a two-dimensional simulation area of linear dimension $L_{sys}$. The migration of the cells is described by the time-dependent position $\vec{r}_{c,n}$ of the respective cell centers, where periodic boundary conditions are applied both in x- and y-direction. Here, $n$ is a discrete time index, related to the continuous time by $t_n = n\;\Delta t_{sim}$.

\vspace{0.1cm}\noindent Throughout this work, we use a fixed simulation time interval of
\begin{equation}
\Delta t_{sim} := 1 \; min.
\label{dtsim}
\end{equation}

\vspace{0.1cm}\noindent The cell trajectories $\vec{r}_{c,n}$ are modelled as discrete time, correlated random walks. In particular, the update from one position to the next is performed as follows:
\begin{equation}
\vec{r}_{c,n} = \vec{r}_{c,n-1} + w_{c,n}\cdot\left(
\begin{array}{c}
     \cos(\;\phi_{c,n-1}+s_{c,n}\;|\Delta\phi_{c,n}|\;)\\
     \,\sin(\;\phi_{c,n-1}+s_{c,n}\;|\Delta\phi_{c,n}|\;)
\end{array}
\right).
\label{updateR}
\end{equation}

\vspace{0.1cm}\noindent In Eq.(\ref{updateR}), $w_{c,n}$ is the step width, which is randomly and independently drawn from a Rayleigh distribution with mean value $v$. Note that this corresponds to an average speed of the cell along the contour of the trajectory (which is a sequence of line segments). 

\vspace{0.1cm}\noindent The quantity $\phi_{c,n-1}$ is the planar angle of motion during the last step of cell $c$, that is, $\phi_{c,n-1}=\arctan(\frac{y_{c,n-1}-y_{c,n-2}}{x_{c,n-1}-x_{c,n-2}})$.

\vspace{0.1cm}\noindent The quantity $\Delta\phi_{c,n}$ is the turning angle between the last and the present step of cell $c$, so that $\phi_{c,n} = \phi_{c,n-1} + \Delta\phi_{c,n}$. The turning angles are randomly and independently drawn from a uniform distribution between the limits $\Delta\phi_{min}(\epsilon)$ and
$\Delta\phi_{max}(\epsilon)$. Here, $\epsilon \in \left[-1,+1\right]$ is a persistence parameter, where $\epsilon=+1$ corresponds to fully persistent motion, $\epsilon=0$ to diffusive motion, and $\epsilon=-1$ to fully anti-persistent motion. Consequently, if $\epsilon>0$, we define $\Delta\phi_{min}(\epsilon)=-(1\!-\!\epsilon)\pi$ and $\Delta\phi_{max}(\epsilon)=+(1\!-\!\epsilon)\pi$. If $\epsilon<0$, we define $\Delta\phi_{min}(\epsilon)=(1\!-\!|\epsilon|)\pi$ and $\Delta\phi_{max}(\epsilon)=(1\!+\!|\epsilon|)\pi$. Note that only the magnitude of the turning angle enters in Eq.(\ref{updateR})

\vspace{0.1cm}\noindent The quantity $s_{c,n}\in\left\{-1,+1\right\}$ is a sign factor, which controls if the cell moves left (counter-clockwise) of right (clock-wise). It is randomly and independently assigned to one of its two possible values, with a probability $prob(R)=prob(s_{c,n} = -\!1)=q_R$. 

\vspace{0.1cm}\noindent The statistical properties of the random walk generated by Eq.(\ref{updateR}) are determined by the three parameters $v$, $\epsilon$, and $q_R$, where $v$ controls the speed of the cells, $\epsilon$ their directional persistence, and $q_R$ their preference to turn left or right (which is usually balanced, so that $q_R=1/2$). In simple cell migration models, these parameters are usually considered as constant over time. However, it has been shown that cell migration is a heterogeneous stochastic process, in which all parameters can change gradually or abruptly, depending on the circumstances of the cell \cite{metzner15,mark18}. In this work, we assume in particular that the immune cell is able to adapt its speed, persistence and left/right preference in response to local gradients of a chemo-attractant.  

\subsection{Assumed size and migration parameters of cells}

If not stated otherwise, simulations in this paper assume that both the immune cell and the target cells are rotation-symmetric and have a radius of
\begin{equation}
r_{imm}=r_{tar} := 10\; \mu m.
\label{rr}
\end{equation}

\vspace{0.1cm}\noindent Target cells are assumed to be slow and to move diffusively:
\begin{equation}
v_{tar} := 1\;\frac{\mu m}{min}; \;\hspace{0.3cm} \epsilon_{tar} := 0.
\label{migTar}
\end{equation}

\vspace{0.1cm}\noindent If required, the immune cell is able to move much faster than the targets and, at least for short periods, with perfect directional persistence:
\begin{equation}
v_{imm} := 0\ldots6\;\frac{\mu m}{min}; \;\hspace{0.3cm} \epsilon_{imm} := 0\ldots1.
\label{migImm}
\end{equation}

\vspace{0.1cm}\noindent Indeed, experiments have shown that natural killer cells can migrate in thin collagen gels with an {\em average} speed of $\overline{v}_{nk}\approx 5.67\;\mu m/min$ and an {\em average} persistence of $\overline{\epsilon}_{nk}\approx 0.64$ [Christoph Mark and Franzsika Hoersch, private communication. Note that CM and FH used a slightly different persistent parameter, which however has the same property that $\epsilon=+1$ corresponds to fully persistent motion, $\epsilon=0$ to diffusive motion, and $\epsilon=-1$ to fully anti-persistent motion].

\subsection{Model for temporal evolution of the chemo-attractant}

Our basic proposition is that the target cells emit a substance into the extra-cellular matrix (mainly consisting of water), which is used as a chemo-attractant by the immune cell. For simplicity, we assume that the chemo-attractant is produced at the center point $\vec{r}_0$ of each target cell with a constant generation rate $g$. The substance is freely diffusing with diffusion constant $D$, and is spontaneously decaying with a rate $k$ (It is important  - and also biologically realistic - that this decay rate is non-zero. Otherwise no stationary density profile will develop). This leads to the following partial differential equation for the time-dependent 2D density distribution of the chemo-attractant $f_{2D}(\vec{r},t)$:
\begin{equation}
\frac{d}{dt} f_{2D} = g\;\delta(\vec{r}-\vec{r}_0) + D\;(\nabla_x^2+\nabla_y^2)\; f_{2D} - k\;f_{2D},
\label{PDE}
\end{equation}

\subsection{Typical parameters of diffusion and decay}

The diffusion constant of a substance within a liquid medium (here basically water) can be estimated by assuming a spherical shape of the diffusing molecules. Using Stokes formula for the friction force, the resulting Stokes-Einstein relation yields
\begin{equation}
 D = \frac{k_BT}{6\pi\eta r},
\end{equation}
where $T=37\;^\circ\text{C}$ is the temperature, $\eta=6.91\cdot10^{-4}\;Pa\,s$ is the  viscosity of water at this temperature, and $r$ is the radius of the diffusing molecule. For a hypothetic molecule with $r=3.18\; nm$, one obtains a diffusion constant of 
\begin{equation}
D := 100\; \mu m^2/s,
\label{dd}
\end{equation}
which will be used throughout this paper. Note that the same value of $D$ was used in an analytical study of the chemo-attractant's density profile \cite{heinrich17}, where the considered molecule was the anaphylatoxin $C5a$. Following this reference, we also assume a typical decay constant of 
\begin{equation}
k := 10^{-2}/s. 
\label{kk}
\end{equation}
The generation rate $g$ is less important in the sense that it does not affect the spatial shape or the temporal evolution of the profile $f_{2D}(\vec{r},t)$.

\vspace{0.1cm}\noindent A dimensional analysis of Eq.(\ref{PDE}) reveals that the system has a characteristic diffusion length of 
\begin{equation}
  L_{dif} = \sqrt{D/k} \approx 100\;\mu m.
  \label{ldif},
\end{equation}
which can be considered as the approximate spatial extent of the density 'cloud' around a stationary emitter. The characteristic time period for developing this density cloud can be estimated as
\begin{equation}
  T_{dif} = \frac{L_{dif}^2}{D} = \frac{1}{k} \approx 100\;s \approx 1.7\; min.
  \label{tdif},
\end{equation}

\subsection{Fast diffusion limit}

Based on the above parameters, we can compute a further characteristic quantity that has the dimensions of a velocity:
\begin{equation}
  v_{crit} = \frac{L_{dif}}{T_{dif}} \approx 60\;\mu m/ min
  \label{vdif},
\end{equation}
If the emitter of the density cloud is moving at a speed much smaller than this critical velocity, we can approximately assume that the density cloud is fully developed at any moment in time. In other words, there will be a cloud of fixed (stationary) shape that is 'carried around' by the emitter along its trajectory. For our assumed typical target cell speed of $v_{tar}=1 \;\mu m/s$, we are indeed well within this 'fast diffusion limit'.

\subsection{Stationary density profile around single target}

The fast diffusion limit saves us from numerically solving the reaction-diffusion equation Eq.(\ref{PDE}). We only need to compute the stationary, rotation-symmetric density profile $f_{2D}(\;r\;=\;|\vec{r}-\vec{r}_0|\;)$ around a non-moving emitter, conveniently located at the origin $\vec{r}_0=\vec{0}$ of the coordinate system. Since the immune cell can never be closer to the emission point $\vec{r}_0$ than the radius $r_{tar}$ of the target cells, we need to solve Eq.(\ref{PDE}) only in the region $r>r_{tar}$, where the generation term disappears. In polar coordinates, these simplifications lead to the following ordinary differential equation for the stationary profile,
\begin{equation}
  \left[\frac{\partial^2}{\partial r^2}\;+\frac{1}{r}\;\frac{\partial}{\partial r}\right]\; f_{2D}(r) = \left(\frac{k}{D}\right) \;f_{2D}(r)
  \label{fr},
\end{equation}
which is solved numerically with a Runge Kutta method. At the border of the prey cell, without restriction of generality, we set the density to $f_{2D}(r=r_{prey})=1$. The slope $\frac{\partial}{\partial r}f_{2D}(r=r_{prey})$ at this point is iteratively adjusted such that $f_{2D}(r\rightarrow\infty)=0$. The resulting radial profile decays rapidly in the direct vicinity of the emitter. For $r\rightarrow\infty$, the decay approaches an exponential shape (see Fig.\ref{First_Panel}(b)).

\subsection{Momentary distribution of chemo-attractant density}

The diffusion profiles of different emitters add up linearly. Therefore, the total distribution of chemo-attractant density from all present target cells can be written as
\begin{equation}
F_{2D}(\vec{r},t=t_n) = \sum_{c=1}^{N_{tar}}f_{2D}(|\vec{r}-\vec{r}_{cn}|).
\end{equation}

\subsection{Modeling sensors for the chemo-attractant}

In the case of {\em temporal sensing}, we assume that the immune cell can measure, in every time step $n$, the total density $\rho^C_n = F_{2D}(\vec{r}=\vec{r}_C,t=t_n)$ of chemo-attractant at the center $\vec{r}_C=\vec{r}_{0,n}$ of its cell body. It then computes the temporal difference
\begin{equation}
\Delta\rho^C_n = \rho^C_n - \rho^C_{n-1}. 
\end{equation}

\vspace{0.1cm}\noindent In the case of {\em spatial sensing}, we assume that the immune cell has two sensors at the left and right border of its cell body, that is, at positions
\begin{equation}
\vec{r}_{L/R}=\vec{r}_{0,n}+r_{imm}\cdot\left(
\begin{array}{c}
\cos(\;\phi_{0,n-1}\pm\pi/2\;)\\
\sin(\;\phi_{0,n-1}\pm\pi/2\;)
\end{array}
\right), 
\end{equation}

\vspace{0.1cm}\noindent where the corresponding total chemo-attractant densities are $\rho^L_n$ and $\rho^R_n$, respectively. It then computes the spatial difference 
\begin{equation}
\Delta\rho^{LR}_n = \rho^R_n - \rho^L_n. 
\end{equation}

\subsection{Mapping sensor signals to migration behavior}

The two sensor signals available to the immune cell are the temporal density difference $\Delta\rho^C_n$ and the spatial density difference $\Delta\rho^{LR}_n$.  The migration parameters which can be affected by these sensor signals are the speed of the immune cell $v$, its directional persistence $\epsilon$, and its preference to turn left $q_L$.

\vspace{0.1cm}\noindent For simplicity, we assume that the immune cell has two distinct migration modes, called the {\em 'normal mode' $N$}, and the {\em 'approach mode' $A$}. In the normal mode, the speed is $v_N$ and the persistence is $\epsilon_N$. In the approach mode, the speed is $v_A$ and the persistence is $\epsilon_A$. These four parameters can be tuned to optimize search performance.

\vspace{0.1cm}\noindent At any time step $n$, the immune cell can only be in one of these two migration modes. The probability to be in the approach mode is computed as a function of the {\em temporal} gradient as follows
\begin{equation}
q_A = prob(A) = logistic(\;c_{A0}+c_{A1}\;\Delta\rho^C_n\;),
\end{equation}
where
\begin{equation}
logistic(x) = \frac{1}{1+e^{-x}}
\end{equation}
is the logistic function, and $c_{A0}$ as well as $c_{A1}$ are unknown coefficients that also have to be optimized. Note that for $c_{A1}>0$, the mode $A$ is favoured whenever there is a positive temporal gradient, provided that the magnitude of the bias $c_{A0}$ is note too large.

\vspace{0.1cm}\noindent In a similar way, the {\em spatial} gradient determines the probability $q_R$ of the immune cell to turn right:
\begin{equation}
q_R = prob(R) = logistic(c_{R1}\;\Delta\rho^{LR}_n\;),
\end{equation}
where $c_{R1}$ is an additional coefficient to be optimized. Note that for $c_{R1}>0$, right turns are favored whenever the chemo-attractant density at the right sensor is larger than that on the left sensor. 

\subsection{Choice of target cell density and linear system size}

The density of target cells in the two-dimensional simulation plane is chosen to be
\begin{equation}
\rho_{tar} := 1\cdot 10^{-5} \mu m^{-2}.
\end{equation}
This density leads to a mean distance between nearest neighbors of
\begin{equation}
\overline{r}_{nn} = \frac{1}{2\sqrt{\rho_{tar}}} \approx 158\; \mu m, 
\end{equation}
which is slightly larger than the diffusion length $L_{dif}=100 \;\mu m$.

\vspace{0.1cm}\noindent The linear system size is chosen as 
\begin{equation}
L_{sys} = 1000\;\mu m,
\end{equation}
which is considerably larger than $L_{dif}$ and $\overline{r}_{nn}$. The average number of target cells within the simulation area is
\begin{equation}
N_{tar} = \rho_{tar}\;L^2_{sys} = 10.
\end{equation}

\vspace{0.1cm}\noindent 
Note that if an immune cell is migrating with its maximum speed of 6 $\mu m/$min and with perfect directional persistence, it would take about 26 min ( = 26 simulation time steps) to cover the distance between two neighboring target cells. Within 100 min, an immune cell of perfect efficiency might encounter 3 to 4 target cells (ignoring the fact that $r_{nn}$ is increasing slightly with each encounter and the simultaneous removal of the target). 

\subsection{Measuring search efficiency}

We thus set the time period of a single simulation run to
\begin{equation}
 T_{sim} := 100 \; min.
\end{equation}

\vspace{0.1cm}\noindent After a specific simulation run $k$, the number of remaining target cells $N_{tar,k}^{rem}$ is counted. We then quantify the efficiency of the immune cell by the number of eliminated target cells:
\begin{equation}
 Q_k = N_{tar}-N_{tar,k}^{rem},
\end{equation}
a quantity that can fluctuate considerable between each run. To overcome these fluctuations, the simulation is repeated
\begin{equation}
 N_{runs} := 10^4
\end{equation}
times for each set of system parameters, using in each run a random initial configuration of the single immune cell and of the $N_{tar}=10$ target cells.

\vspace{0.1cm}\noindent Finally, the {\em search efficiency} of the immune cell is defined as the average
\begin{equation}
 Q = \frac{1}{N_{runs}} \sum_{k=1}^{N_{runs}} Q_k.
\end{equation}

\subsection{Optimization of response parameters
\label{evoOpt}}

In general, the search efficiency in our model depends on up to $n=7$ unknown parameters $\pi_i$:
\begin{equation}
 Q = Q(\vec{\pi}) = Q(v_N, \epsilon_N, v_A, \epsilon_A, c_{A0}, c_{A1}, c_{R1}).
\end{equation}

\vspace{0.2cm}\noindent
Finding the search strategy with the best search efficiency amounts to finding the parameter combination $\vec{\pi}$ that maximizes $Q$:
\begin{equation}
\vec{\pi}_{opt} = arg max \left\{ Q(\vec{\pi}) \right\}
\end{equation}
We perform this quite high-dimensional numerical optimization using a grid-based variant of the 'Cyclic Coordinate Descent' method (CCD, see \cite{wright15}). In each loop of this iterative method, the $n$ parameters/coordinates $\pi_{i\in[1\ldots n]}$ are optimized one after the other in a cyclic way, greedily keeping the remaining $n-1$ coordinates at their presently best-performing values. In our variant of the method, an individual parameter $\pi_k$ is optimized by evaluating $Q=Q(\pi_k,\left\{\pi_{i\neq k}\right\})$ for all discrete values of $\pi_k$ on a regular grid within predefined minimum and maximum values, that is $\pi_k\in[\pi_{k,min},\; \pi_{k,min}\!+\!\Delta\pi_k,\;\ldots\;,\;\pi_{k,max}]$ The method stops when the same set of $n$ optimal parameters is found in two subsequent iteration loops. 

\subsection{List of standard parameters\label{standPar}}

In Tab.(\ref{SP_list}), we provide a list of all relevant system parameters, here called the Standard Parameters (SP). The first 12 parameters of the list are fixed for all simulations. During the optimization phase, a different random value of $v_{tar}$ and $\epsilon_{tar}$ is drawn for each target cell, from uniform distributions in their respective ranges. During the evaluation phase, all $N_{tar}$ target cells are set to the same values of $v_{tar}$ and $\epsilon_{tar}$, and these two parameters are then scanned through their ranges in subsequent simulation runs. The last 7 parameters of the list ($v_N, \epsilon_N, v_A, \epsilon_A, c_{A0}, c_{A1}, c_{R1}$) are free to be optimized within the given ranges.

\section{Results}

\subsection{Concentration profile of chemo-attractant}

\begin{figure}[h!]
\centering
\includegraphics[width=14cm]{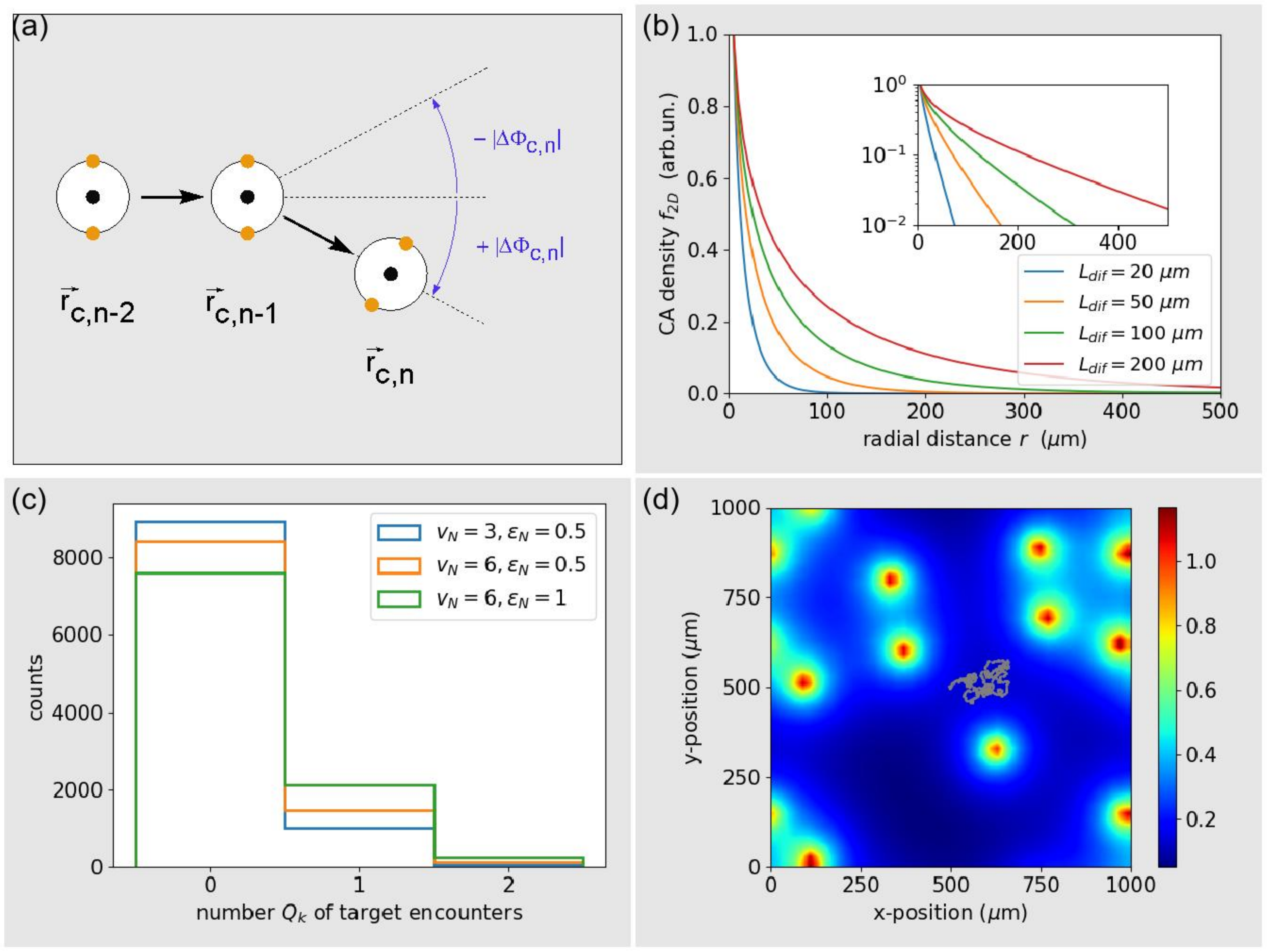}
\caption{
\textit{
(a) Three subsequent positions of the model immune cell (white circles), which is equipped with a central concentration sensor for temporal gradient sensing (black dot) and two lateral concentration sensors for spatial gradient sensing (orange dots). The magnitude of the turning angle $|\Delta\phi_{c,n}|$ can be applied with negative of positive sign (blue).
(b) Stationary radial profile of chemo-attractant density $f_{2D}(r)$ around a non-moving emitter, for different diffusion lengths $l_{dif}$. The semi-logarithmic inset shows that the profile decays almost exponentially for large radial distances $r\rightarrow\infty
$.
(c) Distribution of the number of targets encountered by the immune cell over $10^5$ simulation runs. The three shown cases correspond to the standard parameters (SP, blue), to standard parameters with the immune cell persistence increased to $\epsilon_N=1$ (olive), and to standard parameters with both $\epsilon_N=1$ and speed increased to $v_N=6\; \mu$m/min (red).
(d) Example configuration of static targets (orange dots), concentration distribution of the guiding substance (color code), and the trajectory of the immune cell (small gray dots) over 500 min. The immune cell is set to standard parameters ($c_{A0}=-5$, $c_{A1
}=c_{R1}=0$, $v_N=v_A=3$, and $\epsilon_N=\epsilon_A=0.5$).
}}\label{panel_one}
\end{figure}

In the fast diffusion limit, the global concentration distribution of chemo-attractant is a linear superposition of 'kernels', centered around the target cells. These kernels are the temporally stationary, rotationally symmetric solutions $f_{2D}(r)$ of Eq.(\ref{fr}). We have numerically computed the kernel for different diffusion lengths $L_{dif}$ (See Fig.\ref{panel_one}(b), in which the the green line corresponds to the case of standard parameters). As expected, the concentration profile decays almost exponentially for large radial distances $r\rightarrow\infty$.

\subsection{Blind search (BLS)}

We start with an immune cell that completely lacks the ability to sense concentration gradients ($c_{A1}\!=\!c_{R1}\!=\!0$), and which is therefore performing a 'blind' search process (BLS). At the same time, we assume an extreme bias for the normal migration mode ($c_{A0}=-5$), which reduces the probability of the immune cell being spontaneously in the approach mode to an almost negligible value of $q_A\approx0.007$ (The migration parameters of the approach mode are set to medium values $v_A=3$ and $\epsilon_A=0.5$). The migration of such an immune cell can therefore be described as a homogeneous, correlated random walk with a fixed speed $v_N$ and a fixed degree of directional persistence $\epsilon_N$.

\vspace{0.1cm}\noindent
The $N_{tar}=10$ target cells, which are assigned random positions and migration directions before each simulation run, are assumed to form a widely mixed ensemble with respect to their migration parameters. For this purpose, at the beginning of each simulation run, we draw the speed and persistence parameters of each target cell independently from uniform distributions in the ranges $v_{tar}\in\left[0,6\right]$ and $\epsilon_{tar}\in\left[0,1\right]$, respectively.

\vspace{0.1cm}\noindent
We first set the migration parameters of the immune cell to medium values $v_N=3$ and $\epsilon_N=0.5$. In this case, the trajectory of the immune cell is not able to explore a significant part of the simulation area, even when the available time span is increased from the standard setting $T_{sim}=100$ min to $T_{sim}=500$ min (See Fig.\ref{panel_one}(d). For a video, see \cite{videos}(UBS.mp4)). Repeating the simulation $N_{run}=10^4$ times, each spanning an evaluation period of $T_{sim}=100$, we find that the number $Q_k$ of encounters between the immune cell and target cells is fluctuating from one run $k$ to the next. The distribution $p(Q_k)$ has an approximately exponential shape: In most simulation runs, the immune cell does not find any target, rarely one target, and almost never two targets. The average number of encounters with target cells, defined above as the search efficiency, is $Q=0.110$ in this case. If we let the immune cell migrate faster, using the parameters $v_N=6$ and $\epsilon_N=0.5$, the search efficiency increases to $Q=0.173$. Additionally making the immune cell more directionally persistent, using the parameters $v_N=6$ and $\epsilon_N=1$, results in a further increase of the search efficiency to $Q=0.271$. This demonstrates that even a blind, homogeneous search process can be optimized via the migration parameters $v_N$ and $\epsilon_N$. 

\begin{figure}[h!]
\centering
\includegraphics[width=14cm]{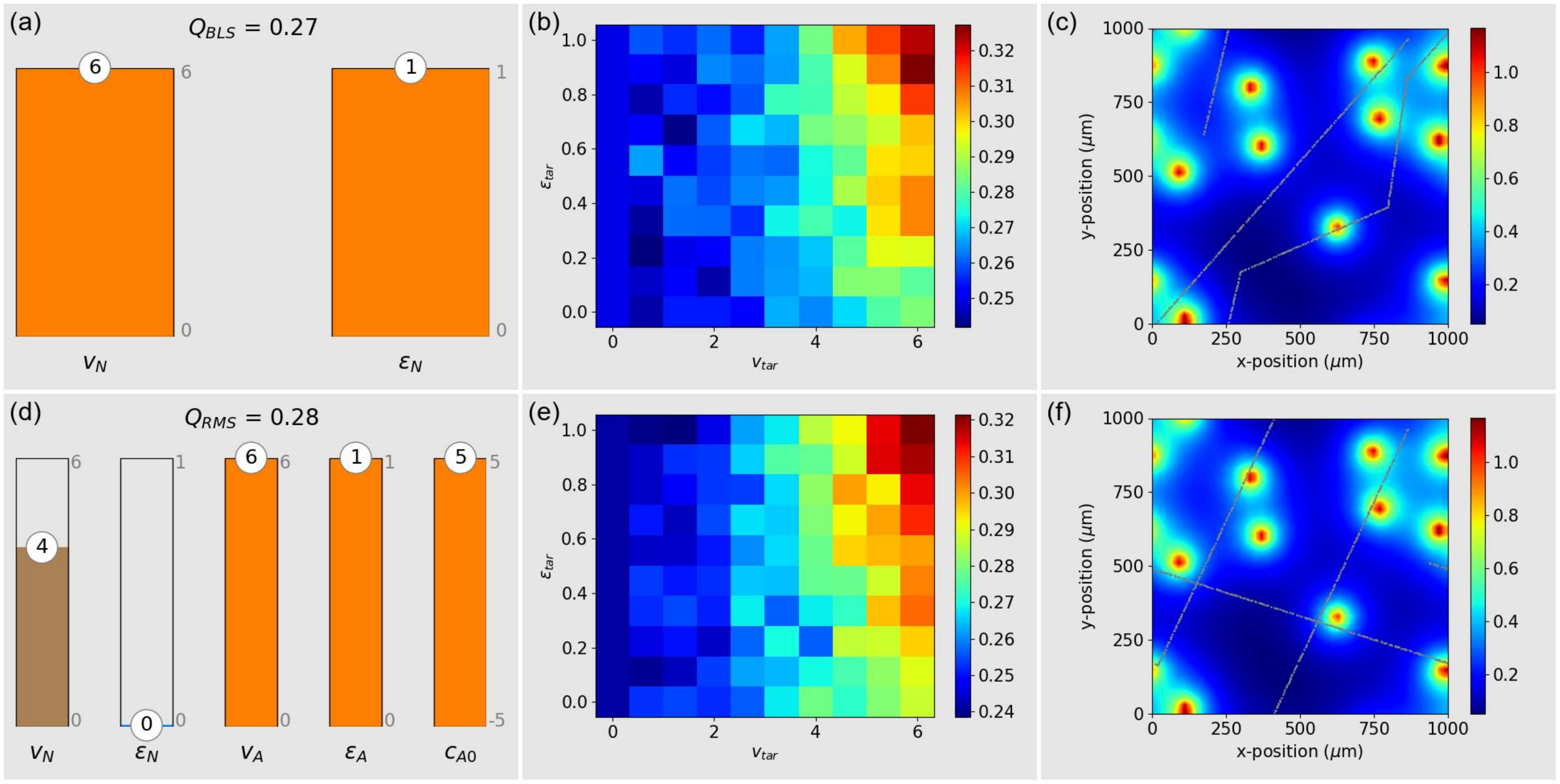}
\caption{
\textit{
{\bf Columns:} Left: Optimal immune cell parameters found by the CCD method. Middle: Search efficiency of the optimized immune cell as a function of the speed ($v_{tar}$) and directional persistence ($\epsilon_{tar}$) of the target cells. Right: Example configuration of static targets (orange dots), concentration distribution of the guiding substance (color code), and the trajectory of the immune cell (small gray dots) over 500 min. {\bf Rows} correspond to different search strategies: Blind search (BLS, top row), and random mode switching (RMS, bottom row). 
}}\label{panel_bls_rms}
\end{figure}

\vspace{0.1cm}\noindent
We therefore use CCD optimization to find the perfect migration parameters for the immune cell, again using the mixed ensemble of target cells throughout the optimization phase. It turns out that a {\em blind, homogeneous search within a mixed ensemble of targets has the best efficiency $Q$ when it is performed with maximum possible speed (in our case $v_N=6$) and with perfect directional persistence $\epsilon_N=1$} (Fig.\ref{panel_bls_rms}(a)).

\vspace{0.1cm}\noindent
The resulting optimal efficiency $Q_{BLS}=0.27$ can be seen as the overall performance of the immune cell, averaged over many possible types of target cells. In practice, it will also be of interest how the immune cell is performing against targets with specific, fixed  migration parameters. To investigate this 'versatility' of the immune cell, we have computed the search efficiency $Q=Q(v_{tar},\epsilon_{tar})$ of the optimized immune cell (that is, using $v_N=6$ and $\epsilon_N=1$), as a function of the speed and persistence of the {\em target} cell (Fig.\ref{panel_bls_rms}(b)). Here we find that the resulting search efficiency can vary between $Q_{min}\approx0.25$ and $Q_{max}\approx0.32$, depending on these two parameters. In particular, {\em blind, homogeneous search works best when the targets are themselves fast and directionally persistent}. Yet, if the targets exceed the immune cell with respect to the migration parameters, it is more appropriate to say that the targets are finding the immune cell than vice versa.

\vspace{0.1cm}\noindent
It is instructive to inspect the trajectory of the immune cell (Small gray dots in Fig.\ref{panel_bls_rms}(c). For a video, see \cite{videos}), in relation to the targets, over an extended time period. For this purpose, we set the speed of the targets to zero, so that they remain stationary throughout the entire simulation. Since the persistence of the optimized immune cell is $\epsilon_N=1$ in the normal mode, the trajectory is straight for most of the time (Note the effect of periodic boundary conditions). However, with a tiny probability of $q_A\approx0.007$, the immune cell also adopts the 'approach mode', where the migration parameters are $v_A=3$ and $\epsilon_A=0.5$, and these rare events lead to an abrupt change of direction. It is remarkable that there occur several 'near misses' between the immune cell and one of the targets. Yet, without any sensing abilities, the immune cell most of the time cannot seize these opportunities.

\subsection{Random mode switching (RMS)}

We continue to consider blind search, characterized by the absence of sensitivity for concentration gradients  ($c_{A1}\!=\!c_{R1}\!=\!0$). But this time we allow the immune cell to switch between its two migration modes randomly and spontaneously, a situation that creates a heterogeneous correlated random walk. For this purpose, we now declare not only the parameters $v_N$ and $\epsilon_N$, but also $c_{A0}$, $v_A$ and $\epsilon_A$ as free, optimizable parameters.

\vspace{0.1cm}\noindent
Although the system is now considerably more flexible than in the case of homogeneous blind search, CCD optimization shows that this flexibility brings no significant improvement of the the search efficiency (Fig.\ref{panel_bls_rms}(d)), as $Q_{RMS}=0.28\approx0.27= Q_{BLS}$. Indeed, the optimal efficiency is found for a bias $c_{A0}=5$, which keeps the immune cell in the approach mode virtually all the time, thus leaving the values $v_N=4$ and $\epsilon_N=0$ irrelevant. Within the approach mode, the optimized immune cell is as fast ($v_A=6$) and persistent ($\epsilon_A=1$) as possible, just like in the above  homogeneous BLS case. This demonstrates that {\em in blind search, purely spontaneous mode switching performs worse than a homogeneous random walk at maximum speed and perfect directional persistence}. Since the optimal RMS strategy is - except for a name change of the dominating migration mode - identical to the BLS strategy, we also find the same results for $Q=Q(v_{tar},\epsilon_{tar})$ (Fig.\ref{panel_bls_rms}(e)). The sample trajectory of the immune cell also resembles that of the BLS strategy (Fig.\ref{panel_bls_rms}(f). For a video, see \cite{videos}).

\subsection{Temporal gradient sensing (TGS)}

Next, we investigate how the killing efficiency can be enhanced when the immune cell is able to measure temporal gradients of the chemo-attractant and to switch between the normal mode $N$ and the approach mode $A$ accordingly. In order to make this adaptive mechanism work, there are six parameters to be optimized: The speeds ($v_N$,$v_A$) and persistence values ($\epsilon_N$,$\epsilon_A$) in the two migration modes, as well as the bias of the approach mode ($c_{A0}$) and the sensitivity for temporal chemo-attractant gradients ($c_{A1}$). Without restriction of generality, the latter quantity is assumed to be non-negative, $c_{A1} \ge 0$, because a positive temporal gradient of the chemo-attractant $\Delta\rho^C_n$ means that the immune cell is approaching a target cell, and this should increase the probability of the approach mode $A$, whatever this means for the speed and persistence of the immune cell.

\begin{figure}[h!]
\centering
\includegraphics[width=14cm]{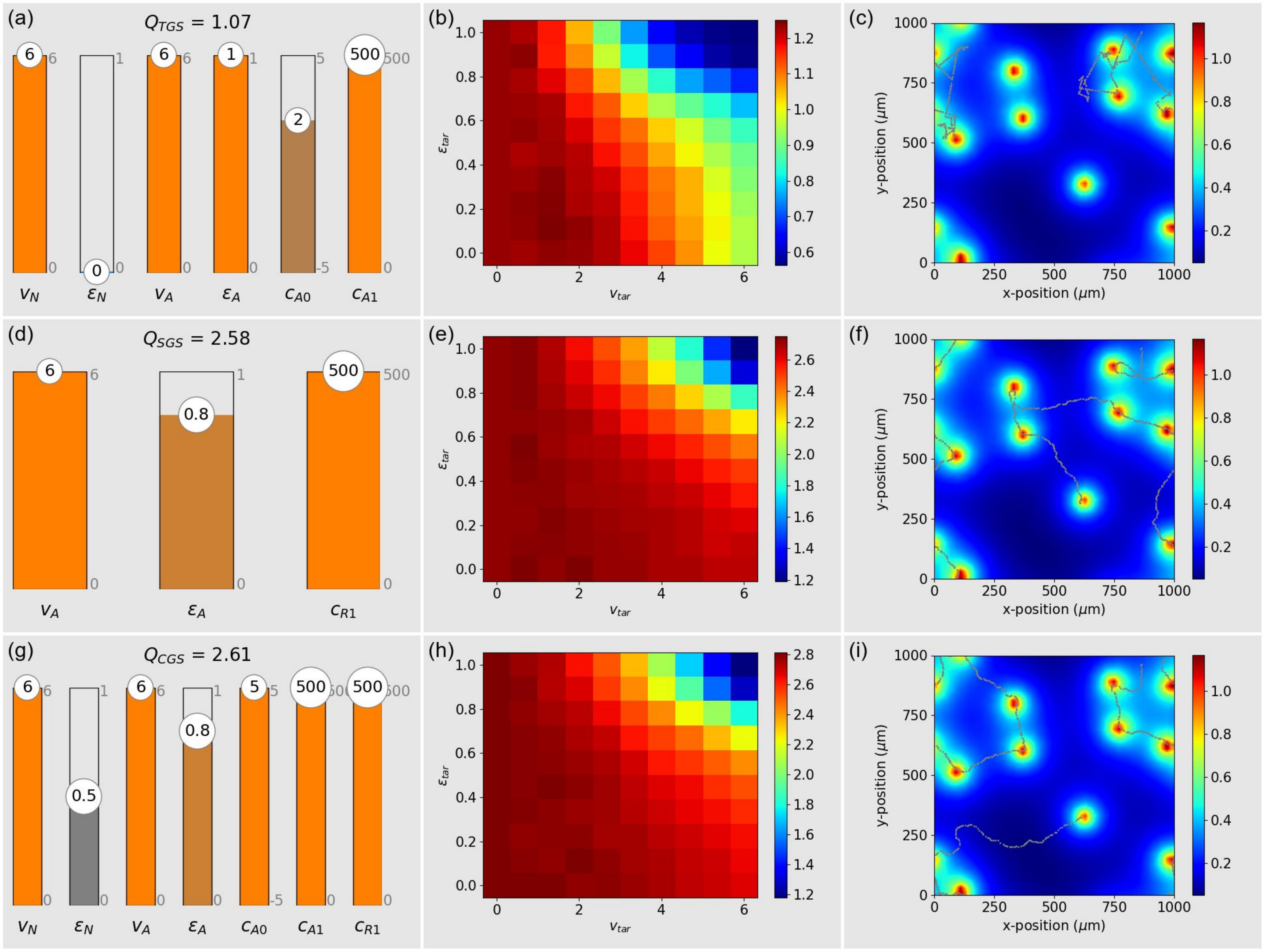}
\caption{
\textit{
{\bf Columns:} Left: Optimal immune cell parameters found by the CCD method. Middle: Search efficiency of the optimized immune cell as a function of the speed ($v_{tar}$) and directional persistence ($\epsilon_{tar}$) of the target cells. Right: Example configuration of static targets (orange dots), concentration distribution of the guiding substance (color code), and the trajectory of the immune cell (small gray dots) over 500 min. {\bf Rows} correspond to different search strategies: Temporal gradient search (TGS, top row), spatial gradient search (SGS, middle row), and combined gradient search (CGS, bottom row).
}}\label{panel_tgs_sgs_cgs}
\end{figure}

\vspace{0.1cm}\noindent
CCD optimization shows (Fig.\ref{panel_tgs_sgs_cgs}(a)) that the optimum bias for the approach mode is $c_{A0}=2$, which corresponds to a probability $q_A\approx0.88$ of the immune cell being in the approach mode if detecting no or only a very weak temporal gradient. When however a significant gradient is present, the large sensitivity parameter $c_{A1}=500$ causes an almost deterministic mode switching behavior: {\em In positive gradients, the optimal TGS cell is adopting the approach mode, which is maximally fast ($v_A=6$) and persistent ($\epsilon_A=1$). In negative gradients, it is adopting the normal mode, which is also fast ($v_N=6$), but directionally non-persistent ($\epsilon_N=0$)}. The resulting search efficiency of the optimized TGS strategy against target cells with mixed migration properties is $Q_{TGS}=1.07$, which surpasses the blind strategies by a factor of $Q_{TGS}/Q_{BLS}\approx 4$. 

\vspace{0.1cm}\noindent
Confronted with target cells of fixed migration properties (Fig.\ref{panel_tgs_sgs_cgs}(b)), the performance of the optimized TGS strategy is degrading relatively quickly when the targets are fast and directionally persistent. 

\vspace{0.1cm}\noindent
The sample trajectory of the immune cell (Fig.\ref{panel_tgs_sgs_cgs}(c). For a video, see \cite{videos}) demonstrates the alternating phases of zero persistence ($\epsilon_N=0$, 'zigzag'-like motion) and perfect persistence ($\epsilon_A=1$, straight motion). In contrast to the blind search strategies, the immune cell is now able to perfectly home in on a target, once it came close to it.  

\subsection{Spatial gradient sensing (SGS)}

We now consider an immune cell that is virtually always in the approach mode (enforced by $c_{A0}=500$), but has the ability to turn left (clock-wise) or right in response to the spatial gradient of the chemo-attractant. The relevant response coefficient for this mechanism is the sensitivity $c_{R1}$. Yet, how well the immune cell can follow a spatial gradient also depends on the migration parameters $v_A$ and $\epsilon_A$, because they determine how quickly the cell can adjust its direction as it follows a spatial gradient. 

\vspace{0.1cm}\noindent
CCD optimization shows (Fig.\ref{panel_tgs_sgs_cgs}(d)) that {\em the optimized SGS immune cell turns into the direction of larger chemo-attractant concentration with maximum sensitivity ($C_{R1}=500$). It migrates with maximal speed ($v_A=6$), but with a specific degree of persistence that is smaller than one ($\epsilon_A=0.8$).}  The resulting search efficiency of the optimized SGS strategy against target cells with mixed migration properties is $Q_{SGS}=2.58$, which surpasses the TGS strategy by a factor of  $Q_{SGS}/Q_{TGS}\approx 2.4$, and blind strategies by a factor of $Q_{SGS}/Q_{BLS}\approx 9.6$.

\vspace{0.1cm}\noindent
Confronted with target cells of fixed migration properties (Fig.\ref{panel_tgs_sgs_cgs}(e)), the optimized SGS cell has a relatively constant performance for targets with small to medium speeds and persistences. In the extreme case of targets with $v_{tar}\approx 6$ and $\epsilon_{tar}\approx 1$, the performance declines, but even then it is still about as good as the optimal TGS performance. 

\vspace{0.1cm}\noindent
The sample trajectory (Fig.\ref{panel_tgs_sgs_cgs}(f). For a video, see \cite{videos}) shows that the optimized SGS immune cell is wasting almost no time between subsequent target attacks. It moves from one target to the next in an efficient way, resembling the optimal solutions of a traveling salesman problem.

\subsection{Combined spatial and temporal gradient sensing (CGS)}

Finally, we consider an immune cell that can, both, switch between two migration modes in response to the temporal chemo-attractant gradient, and at the same time turn left and right in response to the spatial gradient. Since these two mechanisms have different requirements with respect to the migration parameters (For example, temporal sensing requires $\epsilon_N=0$, but spatial sensing works best with $\epsilon_N=0.8$), it is not clear whether a combination of the two abilities is advantageous or reduces the killing efficiency.

\vspace{0.1cm}\noindent
CCD optimization of combined gradient sensing involves the complete set of eight free parameters (Fig.\ref{panel_tgs_sgs_cgs}(g)). The resulting bias $c_{A0}=5$ means that the optimized CGS cell is adopting the approach mode practically all the time. In this mode, it just performs spatial gradient sensing, since all the parameters that are relevant to SGS are actually unchanged ($c_{R1}=500$, $v_A=6$, and $\epsilon_A=0.8$). However, the optimized CGS cell is also highly sensitive to temporal gradients ($c_{A1}=500$). Therefore, in the presence of a sufficiently negative temporal gradient, it will switch to the normal mode, which is fast ($v_N=6$) but only medium persistent ($\epsilon_N=0.5$). This means that {\em Combined gradient sensing is basically like spatial gradient sensing, but with the additional feature of a less persistent migration in strongly negative temporal gradients}. The resulting search efficiency of the optimized CGS strategy against target cells with mixed migration properties is $Q_{CGS}=2.61$, which is only slightly better than the SGS strategy. The versatility of combined gradient sensing resembles that of purely spatial gradient sensing (Fig.\ref{panel_tgs_sgs_cgs}(h)). Also the sample trajectory (Fig.\ref{panel_tgs_sgs_cgs}(i). For a video, see \cite{videos}) has basically the same characteristics as in the SGS strategy.

\section{Discussion and Summary}

Chemotaxis is an important phenomenon in prokaryotic cells, such as bacteria \cite{adler66,eisenbach04}, but it plays an equally fundamental role in eukaryotes \cite{swaney10}. Providing one of the simplest examples of goal-directed behavior, chemotaxis is also a fascinating topic that has attracted the interest of physicists since decades.
Among the physics-oriented publications, many focus on the formation of spatio-temporal patterns in colonies of self-propelled agents with mutual predator-prey relations \cite{keller71, czirok96, boraas98, tsyganov03, pang04}. While most of these studies describe the complete density distribution of the agents by partial differential equations, as in the Keller-Segel model, a few works also describe the motion of the agents individually, for example in the framework of active Brownian particles \cite{romanczuk08}. A more recent theoretical study \cite{sengupta11} investigates the chemotactic pursuit of a single prey agent by a predator. Although this work addresses a research question similar to ours, it is based on different model assumptions. In particular, it assumes not only that the predator is chemically attracted by the prey, but also that the prey is repelled from the predator. Furthermore, the guiding chemicals in \cite{sengupta11} are assumed to have an infinite life time, which prevents the formation of a stable chemical 'cloud' around each agent and leads to long-range interactions.

Most existing physics-oriented models of single-agent chemotaxis describe the motion of the predator by a Langevin equation of an over-damped particle that feels a delta-correlated random force, plus a deterministic force proportional to the local concentration gradient of the guiding substance. According to this model, in a situation with zero gradient, the predator will perform a Gaussian random walk without any directional correlations. In the presence of a  non-zero gradient, a force will act on the predator that pulls it into the direction of the prey.

In reality, however, cells typically move with a high degree of directional persistence, even without detecting any chemical concentration differences. Also, changing the migration direction of a cell is a complex process that involves, among many other stochastic events, a partial remodeling of the cytoskeleton. Even though a concentration gradient is pointing strongly into a particular direction, the predator may not be able to turn into this new direction immediately and with sufficient precision. Instead of describing chemotaxis as a deterministic force, an indirect, stochastic approach may therefore be more appropriate: The migration of the predator is a stochastic process, and the parameters of this process (such as the probability of turning left or right) are modified as a function of the chemotactic gradient.

Apart from this opportunity to improve the modeling of chemotaxis, the present work was motivated by preliminary experiments \cite{PrivCom} on chemotactic 'pursuit' in a Petri dish. These experiments studied the interaction of natural killer (NK) cells \cite{vivier08}, extracted from the blood of human donors, and human leukemic K562 tumor cells. By following the migration path of all cells over several hours, it was observed that a fraction of the highly mobile NK cells approached individual tumor cells (which themselves showed only weak mobility) in a directionally persistent walk, attacked them, and often induced their death subsequently. These in-vitro experiments demonstrated that single immune cells are able to find certain pathogenic targets on their own account. However, it remained unclear if the observed attacks were merely chance encounters, or actually guided by chemotactic mechanisms.

In order to answer this question, it would certainly be useful to know which efficiencies can be expected from a 'blind' search, and from different 'guided' search strategies based on chemotaxis. In addition, finding distinct, highly efficient and robust search strategies by numerical optimization of a simulated immune cell would also reveal characteristic patterns of search behavior, that might then be used as characteristic 'fingerprints' of goal-directed search in future automatic detection algorithms. 

For these reasons, we have compared in this work five distinct strategies of search, namely blind search with fixed speed and directional persistence of the immune cells (BLS), blind search with random switching between two distinct migration modes (RMS), guided search based on temporal gradients of the chemo-attractant (TGS), guided search based on spatial gradients of the chemo-attractant (SGS), and a combination of temporal and spatial sensing (CGS). Throughout our study, we have kept the system geometry (two dimensions, as on a Petri dish) and all parameters (density of the target cells, properties of the chemo-attractant, sizes and migration properties of the cells, sensing abilities of the immune cell) close to experimentally realistic values.

In the case of blind search (BLS), not surprisingly, the search efficiency of the immune is almost an order of magnitude lower than with the best guided search mechanisms. Nevertheless, since many pathogens will not emit any chemical substance that the immune cell can detect and use as a guide to its target, blind search may often be the only option. It is therefore fortunate that blind search can be easily optimized by making the immune cell as fast and directionally persistent as possible. This can be understood most easily assuming immobile target cells that are located at random positions within the plane. As the search time $t$ is going on, the blindly migrating immune cell is exploring more and more regions of the Petri dish, and we can mentally mark all spatial pixels that have been visited at least once by the immune cell. The total area of all marked pixels, $A(t)$, here called the 'visited area', is growing monotonously with time, and all target cells that happen to be located within the visited area can be considered as found by the immune cell. Their expected number is $\overline{N}_{found}(t)=\rho_{tar}\;A(t)$, where $\rho_{tar}$ is the areal density of target cells. If the immune cell is migrating with low directional persistence, it will re-visit many pixels more often than once, which is counter-productive with target cells that never move. In this case, the visited area will grow sub-linearly with time. By contrast, $A(t)\propto t$ for an immune cell that is migrating with perfect directional persistence and constant speed, that is, uniformly along a straight line. It is therefore clear that high directional persistence is an important way to improve the blind search efficiency $Q$ of immune cells. At the same time, speed is another key factor for efficient search: For an immune cell in uniform motion, the expected number of found target cells at the end of the search period, $\overline{N}_{found}(t=T_{sim})\propto v_{imm}$, will be directly proportional to its speed.

However, it is well-known that actual cells - and not only immune cells - are showing gradual or abrupt changes of their speed and persistence \cite{metzner15,mark18}, so that their migration has to be described by a temporally heterogeneous stochastic process. The result of such parameter fluctuations are often 'anomalous' properties of the cell's random walk, such as a mean squared displacement that increases with lag-time approximately as a powerlaw. It is not clear whether  temporally heterogeneous cell migration is just a side effect of other causes (such as differences in the local micro-environment of the migrating cell or internal changes connected with the cell cycle), or if it actually serves a purpose. Theoretically, the heterogeneity may help to increase the blind search efficiency of an immune cell, particularly when the targets are mobile. We have therefore investigated how the search efficiency is affected when the immune cell performs random switches between two different migration modes (the RMS strategy). Yet, as suggested by the theoretical argument above, the RMS strategy did not perform significantly better than blind search with fixed migration parameters.

Next, we have investigated guided search strategies that are based on the sensing of chemotactic gradients. In the case of temporal gradient sensing (TGS), we found that the optimized immune cell is switching between two distinct migration modes: In positive gradients, it is adopting the approach mode, which is maximally fast ($v_A=6$) and persistent ($\epsilon_A=1$). By this way, the cell is climbing up the gradient consistently, which usually corresponds to approaching one of the targets. In negative gradients, it is adopting the normal mode, which is also fast ($v_N=6$), but directionally non-persistent ($\epsilon_N=0$). In this mode, the cell is exploring new migration directions, until it finds one with a positive gradient. Note that the optimal TGS strategy found here by numerical parameter optimization strongly resembles the chemotaxis behavior of {\em Escherichia Coli} \cite{neidhardt87}, with its gradient-dependent switching between swimming and tumbling modes of migration. Compared to blind search, TGS is more effective. On the other hand, the gained factor of four in search efficiency is not really large.

In the case of spatial gradient sensing (SGS), we found that the optimized immune cell turns into the direction of larger chemo-attractant concentration with maximum sensitivity ($C_{R1}=500$). It migrates with maximal speed ($v_A=6$), but with a specific degree of persistence that is smaller than one ($\epsilon_A=0.8$). Presumably, this specific degree of persistence represents an optimal compromise between the need to maximize the visited area, and the need to perform clockwise and counter-clockwise turns with the right curvature. Compared to blind search, SGS is almost an order of magnitude more efficient. A combination of temporal and spatial sensing (CGS) turned out to bring no significant advantages compared to pure spatial sensing. 

The blind and guided search strategies differ characteristically in how the search efficiency $Q=Q(v_{tar},\epsilon_{tar})$ depends on the migration parameters of the targets: While blind search (BLS, RMS) works better with fast and persistent targets, the opposite is true for guided search (TGS, SGS, CGS). In guided search, due to the optimization against a mixed set of targets, the search efficiency $Q=Q(v_{tar},\epsilon_{tar})$ remains approximately constant for most combinations of $v_{tar}$ and $\epsilon_{tar}$. Only for targets that are simultaneously extremely fast and persistent does $Q$ decline significantly. Assuming an experimental possibility to vary the migration properties of the targets, without affecting the immune cell or the properties of the chemo-attractant, this predicted difference in $Q=Q(v_{tar},\epsilon_{tar})$ offers a first possibility to distinguish between blind and guided search strategies.

Finally, our work suggests how to detect different search strategies of an immune cell by looking for characteristic patterns in the cell's trajectory: In the case of temporal sensing, the immune cell will show alternating phases of low and high directional persistence, and the probability of the high persistence mode will increase whenever the immune cell approaches one of the targets. In the case of spatial sensing, the left- and right-turns of the immune cell will occur in such a way that they tend to align the cell into the direction of the closest target.

In future work, our investigation could be improved and extended in several obvious ways. For example, we have so far assumed that the immune cell is able to detect arbitrarily small concentrations (or differences between two concentrations) of the chemo-attractant. A lower detection limit may very well change the optimal search parameters and, accordingly, the associated search strategies. Furthermore, it would be straight-forward to extent our simulations to three spatial dimensions. Since the local concentration is, at least within the fast diffusion limit, just a sum over fixed kernels, a change from 2D to 3D kernels would not increase the computation time. By contrast, going beyond the fast diffusion limit is computationally more demanding, as it requires to solve the partial differential equation of the spreading and decaying chemo-attractant along with the motion of the cells. However, we have already demonstrated the feasibility of this approach in 2D (data not shown).

\clearpage
\section*{Acknowledgements}

We thank Christoph Mark and Franziska Hoersch for many helpful discussions, and for sharing preliminary experimental results on the interaction between natural killer cells and tumor cells in a Petri dish.

\section*{Funding}
This work was funded by the Grant ME1260/11-1 of the German Research Foundation DFG.

\clearpage
\section*{References}

\bibliographystyle{unsrt}
\bibliography{refs}


\clearpage
\begin{table}
\begin{tabular}{| c | r | c | l |}\hline
{\bf Symbol} & {\bf Value} & {\bf Unit} & {\bf Description} \\\hline
&&&\\\hline
$L_{sys}$ & 1000 & $\mu$m & Linear system size \\\hline
$\Delta t_{sim}$ & 1 & min & Simulation time step  \\\hline
$T_{sim}$ & 100 & min & Total simulation time per run \\\hline
$N_{runs}$ & 10000 & - & Number of runs per parameter set \\\hline
&&&\\\hline
$D$ & 100 & $\mu$m$^2$/sec & Diffusion constant of chemo-attractant \\\hline
$k$ & 0.01 & sec & Decay rate of chemo-attractant (CA) \\\hline
&&&\\\hline
$N_{tar}$ & 10 & - & Initial number of target cells \\\hline
$r_{tar}$ & 10 & $\mu$m & Radius of target cells \\\hline
$v_{tar}$ & $\left[0,6\right]$ &  $\mu$m/min & Speed of target cells, uniformly distributed \\\hline
$\epsilon_{tar}$ &  $\left[0,1\right]$ & - & Persistence of target cells, uniformly distributed  \\\hline
&&&\\\hline
$N_{imm}$ & 1 & - & Number of immune cells \\\hline
$r_{imm}$ & 10 & $\mu$m & Radius of immune cell \\\hline
&&&\\\hline
\rowcolor{Gray}
$v_N$ & $\left[0,6\right]$ &  $\mu$m/min & Speed of immune cell in normal mode \\\hline
\rowcolor{Gray}
$v_A$ & $\left[0,6\right]$ &  $\mu$m/min & Speed of immune cell in approach mode \\\hline
\rowcolor{Gray}
$\epsilon_N$ & $\left[0,1\right]$ & - & Persistence of immune cell in normal mode \\\hline
\rowcolor{Gray}
$\epsilon_A$ & $\left[0,1\right]$ & - & Persistence of immune cell in approach mode \\\hline
\rowcolor{Gray}
$c_{A0}$ & $\left[-5,5\right]$ & - & Bias of immune cell for approach mode \\\hline
\rowcolor{Gray}
$c_{A1}$ & $\left[-500,500\right]$ & - & Sensitivity of immune cell for temporal CA differences \\\hline
\rowcolor{Gray}
$c_{R1}$ & $\left[-500,500\right]$ & - & Sensitivity of immune cell for spatial CA differences \\ \hline
\end{tabular}
\caption{
\label{SP_list}  
Table of standard, fixed simulation parameters (white background), and the seven free parameters that can be optimized (gray background). Throughout this paper, we implicitly assume that all fixed parameters are set according to this table.
}
\end{table}
\newpage

\clearpage

\clearpage

\clearpage

\end{document}